\documentclass[12pt,onecolumn]{article}

\usepackage{amsmath,amssymb}
\usepackage{times}
\usepackage{xcolor}
\usepackage{graphicx}
\usepackage[font=small]{caption}
\usepackage{float}
\usepackage[switch]{lineno}
\usepackage[
  backend=biber,
  sorting=none
]{biblatex}
\DeclareBibliographyDriver{article}{%
  \usebibmacro{bibindex}%
  \usebibmacro{begentry}%
  \usebibmacro{author/editor+others/translator+others}%
  \setunit{\labelnamepunct}\newblock
  \usebibmacro{title}%
  \newunit
  \usebibmacro{journal+issuetitle}%
  \newunit
  \usebibmacro{doi+eprint+url}%
  \newunit\newblock
  \usebibmacro{addendum+pubstate}%
  \setunit{\bibpagerefpunct}\newblock
  \usebibmacro{pageref}%
  \usebibmacro{finentry}%
}
\renewbibmacro{in:}{%
  \ifentrytype{article}{}{\printtext{\bibstring{in}\intitlepunct}}%
}
\renewbibmacro*{journal+issuetitle}{%
  \usebibmacro{journal}%
  \setunit*{\addspace}%
  \iffieldundef{series}
    {}
    {\newunit
     \printfield{series}%
     \setunit{\addspace}}%
  \printfield{volume}%
  \setunit{\addcolon\space}%
  \printfield[bold]{number}% ← Bold issue number
  \setunit{\addcomma\space}%
  \printfield{pages}%
  \setunit{\addcomma\space}%
  \iffieldundef{year}
    {}
    {\printtext[parens]{\printdate}}% ← Normal font year in (parentheses)
  \newunit}
\renewbibmacro*{finentry}{\relax}
\DeclareFieldFormat
  [article,inbook,incollection,inproceedings,patent,thesis,unpublished]
  {title}{#1}

\begin{document}

\title{Ultrafast optical gating in a nonlinear lithium niobate microcavity}

\author{Ouri Karni$^{1\ast}$, Chirag Vaswani$^1$, and Thibault Chervy$^{1\ast}$}

\date{\today}
%\twocolumn[
\maketitle

\noindent \normalsize{$^{1}$NTT Research, Inc. Physics \& Informatics Laboratories, 940 Stewart Dr, Sunnyvale, CA 94085}
\begin{center}
\normalsize{$^\ast$ Corresponding authors:\\
Thibault Chervy, thibault.chervy@ntt-research.com; Ouri Karni, ouri.karni@ntt-research.com}
\end{center}
%]
%\textbf{One-sentence summary}. We have found a new type of excitons that exist at the interface between two semiconductors. 
\textbf{Recent advances in optical simulation and computational techniques have renewed interest in high-finesse optical cavities for applications such as enhancing light-matter interactions, engineering complex photonic band structures, and storing quantum information. However, the extended interaction times enabled by these cavities often come at the cost of slow optical read-out protocols and limited control over system transients. To address this challenge, we demonstrate an ultrafast intra-cavity optical gating scheme in a high-finesse, second-order nonlinear microcavity incorporating a thin-film of lithium niobate. A femtosecond optical gate pulse --tuned to the transparency region of the cavity's dielectric mirrors-- achieves instantaneous up-conversion of the intra-cavity field via sum-frequency generation. The resulting upconverted signal exits the cavity as a short pulse, providing space- and time-resolved, on-demand access to the intra-cavity state. We validate this approach by tracking the dynamics of multiple resonant modes excited in a plano-concave distributed Bragg reflector microcavity, showing close agreement with analytical models. Additionally, we demonstrate that stimulated intra-cavity difference-frequency generation can efficiently instantiate cavity modes on femtosecond timescales. This gating scheme is fully compatible with low-temperature microcavity experiments, paving the way for advanced quantum state storage, retrieval, and real-time control of light-matter interactions.}

%\keywords{Microcavity, Thin-film lithium niobate, Ultrafast spectroscopy, Ultrafast imaging, Up-conversion spectroscopy}

% \section*{Introduction}\label{sec1}

Optical cavities play a central role in the development of experimental quantum optics. Confining light to a finite volume of space forces photons to occupy discrete sets of resonant modes, defined by their wavelengths, polarization, and spatial field profiles. These modes circulate in the cavity for extended durations, thus prolonging the light-matter interaction time and enabling new applications in acute sensing and spectroscopy \cite{gagliardi2014cavity, BitarafanSensingRev2017} and efficient nonlinear \cite{kippenberg_microresonator-based_2011, mckenna_ultra-low-power_2022} and quantum optics \cite{review_ACSNano2025_Qnonlinearoptics}.

In this context, Fabry-Perot micro-cavities have emerged as versatile platforms for the simulation of solid-state Hamiltonians \cite{PhysRevLett.112.116402, klembt2018exciton, byrnes2014exciton, plumhof_room-temperature_2014, delteil_towards_2019} and many-body systems \cite{kasprzak_boseeinstein_2006,deng_exciton-polariton_2010, tan2023bose}, all-optical information processing \cite{sanvitto2016road, kavokin2022polariton, berloff_realizing_2017} and storage \cite{barland2002cavity, paraiso2010multistability, cerna2013ultrafast}, and are instrumental in the advancement of polaritonic chemistry \cite{ebbesen2016hybrid, garcia2021manipulating, ebbesen2023introduction}. In these systems, the resonant light bouncing between the cavity mirrors dynamically evolves as it interacts with the host material, forming, for example, exciton-polariton states. This evolution occurs over multiple time-scales spanning from the fast round-trips between the mirrors and across the transverse plane, through any material-related dynamics, to the slow leakage of light out of the cavity. Hence, the time-dependent state of light resonating in the cavity is carrying key information, for instance on the converging results of a quantum simulator \cite{minev2019catch}, or on the formation of complex many-body excitations \cite{clark2020observation}, constantly motivating the development of new techniques to instantiate, control, and measure these field transients.

So far, most of the techniques developed for this purpose have relied on the slow decay of the intra-cavity field through the microcavity mirrors, and its characterization by {\it e.g.}, time-resolved photon-counting \cite{Arakawa2018_TR_PL}, up-conversion spectroscopy \cite{Norris_Weisbusch_1994_upconv_interfer,Virgili_upconversiopn_2015, Delteil:19}, spectral interferometry \cite{Norris_Weisbusch_1994_upconv_interfer}, or on collecting indirect observable such as transient absorption signals \cite{deleglise2008reconstruction, Tan2020_pump_probe}. However, to fully exploit the potential of microcavity systems, it is necessary to implement dynamical control schemes where the coupling between the intra-cavity field and its environment can be tuned over time-scales much shorter than cavity mode lifetime. Such dynamical control schemes are particularly important in the context of storage and retrieval of quantum optical information, where the coherent extraction of the intra-cavity state into a single output temporal mode could facilitate quantum state tomography and enable cascaded operations in a well-defined mode basis \cite{reddy2018photonic, heuck2020controlled, heuck2020photon, bustard2022toward}. Despite its great potential, dynamic control of high quality photonic structures remains largely unexplored, and have not been demonstrated in standard Fabry-Perot micro-cavities.

In this work, we introduce a novel platform for ultrafast dynamical control of light by integrating thin-film lithium niobate (TFLN) with a high-finesse, tunable zero-dimensional (0D) microcavity. To demonstrate the platform’s capabilities, we show that the intra-cavity field can be extracted on femtosecond timescales through second-order nonlinear optical gating by an external control pulse. This technique enables coherent mapping of the cavity field's spatiotemporal dynamics with femtosecond temporal and micrometer spatial resolution. Additionally, we demonstrate nonlinear excitation of the cavity field via stimulated intra-cavity difference-frequency generation, enabling ultrafast and efficient instantiation of optical modes in a high finesse resonator. The platform is fully compatible with low-temperature operation, as demonstrated in a closed-loop cryostat at 4\,K, making it suitable for integration with quantum optical and cryogenic photonics experiments.

\begin{figure*}[!t]
\centering\includegraphics[width=\linewidth]{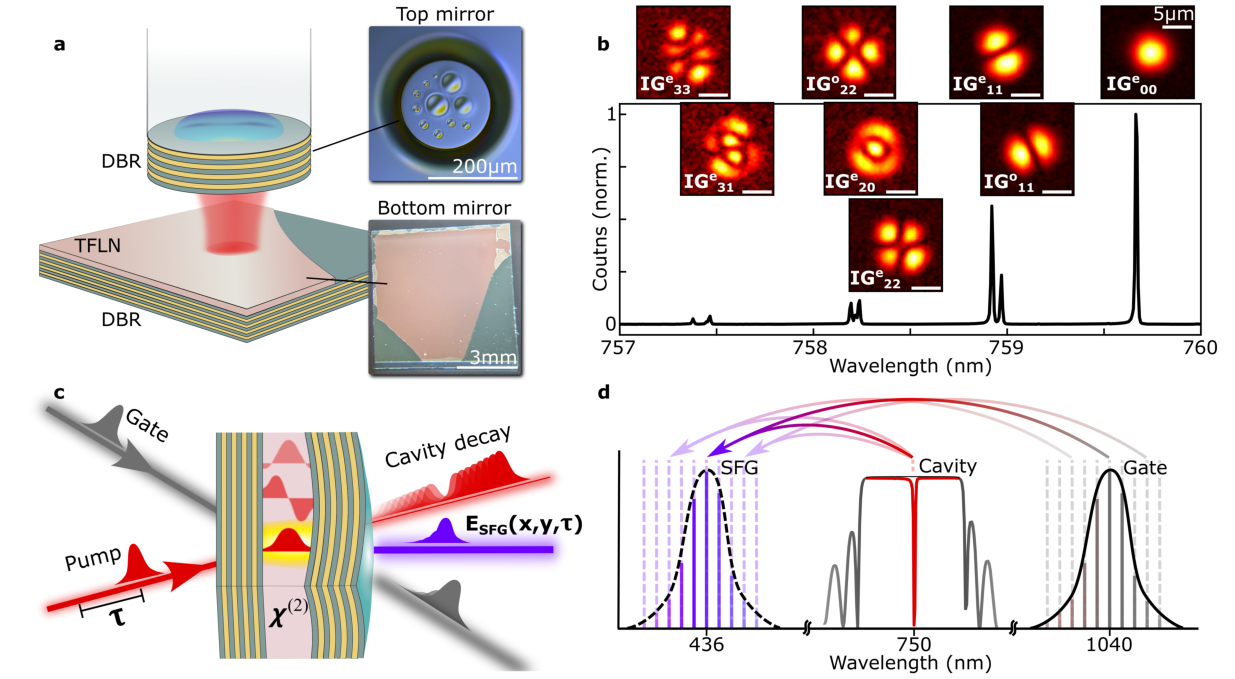}
\caption{\label{fig:Figure_1} \textbf{a.} Schematic of the nonlinear microcavity structure. A $10\,\mu$m thick Mg:LiNbO$_3$ slab (TFLN) is wafer-bonded to the bottom flat DBR mirror (lower inset: micrograph of diced chip). The top curved mirror is held at a controllable distance above, defining the cavity length and transverse mode confinement (top inset: phase-contrast micrograph, featuring several curved micro-mirrors with varied radii of curvature, fabricated on an elevated SiO$_2$ mesa). A radius of curvature of $30\mu\,$m was used in our experiments. \textbf{b.} Transmission spectrum of a {\it ca.} $12\,\mu$m long cavity, illuminated by a broadband source polarized along the extra-ordinary axis of the TFLN slab, showing successive transverse resonances (measured mode images in insets). \textbf{c.} Illustration of the optical gating procedure. Cavity modes are resonantly populated by a 150\,fs pump pulse. The co-propagating gating pulse (angled here is for clarity) arrives at a controlled delay time $\tau$, generating the SFG signal. \textbf{d.} Spectral arrangement of the experiment. The resonant cavity mode near 750\,nm (red) is within the DBR stop-band. The gate pulse (1040\,nm) and SFG signal (436\,nm) are located in the transparency regions of the DBR.}
\end{figure*}

The device at the basis of our experiment is shown schematically in Fig.\,\ref{fig:Figure_1}a. It consists of a plano-concave dielectric Bragg reflectors (DBR) microcavity, embedding a thin film of X-cut MgO:LiNbO$_3$ (TFLN). The $10\,\mu$m thick TFLN layer is wafer-bonded to the flat DBR substrate, and a curved micro-structured DBR mirror ($30\,\mu$m radius of curvature) is held at a controlled distance above by piezo actuators, forming a high finesse tunable microcavity (see Methods section for fabrication details). A representative transmission spectrum of the microcavity, polarized along the extra-ordinary axis of LiNbO$_3$, is shown in Fig\,\ref{fig:Figure_1}b for a fixed cavity length, displaying the expected series of Ince-Gaussian modes \cite{bandres_incegaussian_2004}. Note that the in-plane crystal anisotropy of X-cut TFLN splits the manifold of cavity modes into two orthogonally polarized series. In the following, we fix the polarization of all beams along the crystal extra-ordinary axis, corresponding to the largest $\chi^{(2)}$ coefficient of TFLN.

The gating procedure is schematically presented in Fig.\,\ref{fig:Figure_1}c. Resonant modes of the cavity are first excited by a short optical pulse (150\,fs) centered at 750\,nm wavelength, within the DBR stop-band. After an adjustable delay $\tau$, the gating pulse (150\,fs, $\sim 30\,\mu$m spot diameter) is launched at the cavity. This pulse is centered at 1040\,nm wavelength, outside of the DBR stop-band, and thus passes only once through the cavity, retaining both its short-pulse and wide-spot characters. While traversing the TFLN layer, the gating pulse up-converts the instantaneous and local intra-cavity field into a sum-frequency generated (SFG) signal around 436\,nm. As depicted in Fig.\,\ref{fig:Figure_1}d the SFG spectrum is well outside of the DBR stop-band, and therefore quickly escapes the cavity as a pulse as well. The SFG signal thus extracts instantaneous and local information about the intra-cavity field, which we then measure in two complementary modalities: time-resolved SFG images, $I_{SFG}(x,y,\tau)$, and time-resolved SFG spectrograms $S_{SFG}(\omega,\tau)$ (see Methods section). The results we show below, alongside an analytical model of the gating procedure, serve to explore the capabilities of this technique. 

\section*{Results}
\subsection*{Evolution of individual modes}\label{subsec2_1}

\begin{figure*}[ht]
\centering\includegraphics[width=\linewidth]{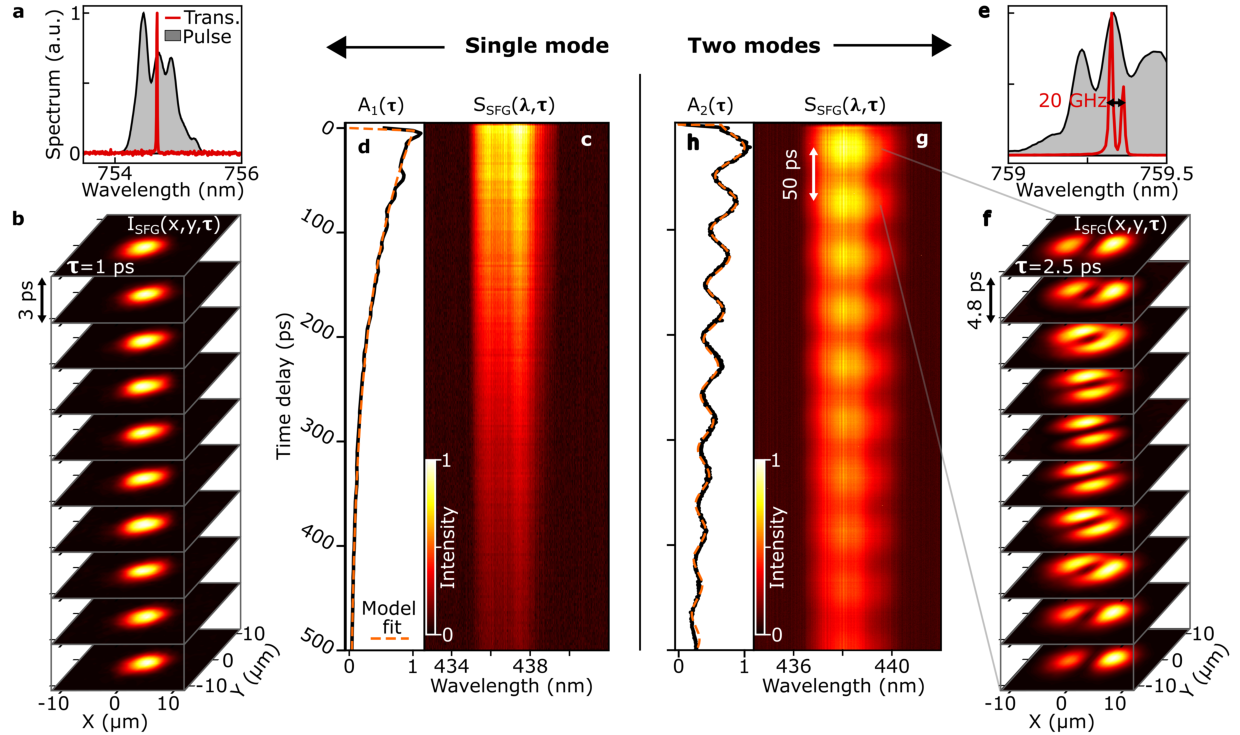}
\caption{\label{fig:Figure_2} \textbf{a, e.} Transmission spectra of the cavity (red) when excited with a filtered pulse (gray) centered on the (a) fundamental and (e) $IG^e_{11}$ and $IG^o_{11}$ cavity modes.  \textbf{b, f.} Snapshots of the transmitted SFG images at specific time-delays. \textbf{c, g.} SFG spectrograms. For clarity, the normalized intensity are corrected by a power of 0.5. \textbf{d, h.} Integral of the SFG spectrograms over the wavelength coordinates (black curve), and fitted decay dynamics (d) $A_1(\tau)$, (h) $A_2(\tau)$ (orange dashes).}
\end{figure*}

We begin by considering two simple-to-interpret scenarios: when the cavity is excited in only a single resonant mode, and then when a pair of modes is excited. Each is achieved by spectral filtering of the excitation pulse around the desired mode spectrum, as shown in Fig.\,\ref{fig:Figure_2}a and e. 

Upon selectively exciting the fundamental $IG^e_{00}$ mode, the time-resolved SFG images present a steady spot profile that resembles the $IG^e_{00}$ distribution, with negligible changes over time, as shown in Fig.\,\ref{fig:Figure_2}b (see also video in Supplementary Material). The SFG spectrogram, recorded over 500\,ps in Fig.\,\ref{fig:Figure_2}c, shows a smooth amplitude decay across the entire SFG spectrum, in accord with the expected dynamics of a single eigenmode excited in the cavity. 

The dynamics become more elaborated when a pair of modes is excited. We specifically choose the pair of $IG^e_{11}$ and $IG^o_{11}$ modes, with a single node in their spatial shape and a 20\,GHz spectral gap between them due to residual astigmatism in the parabolic microcavity mirror (Fig.\,\ref{fig:Figure_2}e). The SFG images now display periodic transitions between two rotated copies of the original Ince-Gaussian mode profiles, as shown in the selected snap-shots in Fig.\,\ref{fig:Figure_2}f (full video in Supplementary Material). These correspond to the symmetric and anti-symmetric superpositions of the $IG^e_{11}$ and $IG^o_{11}$ eigen-modes. The periodicity of these transients is about 50\,ps, in agreement with the inverse of the spectral gap between the modes. The SFG spectrogram (Fig.\,\ref{fig:Figure_2}g) features corresponding temporal beatings with the same periodicity, that show-up uniformly across the SFG spectrum. 

A quantitative understanding of these results can be obtained by comparing the data to an analytical derivation of the transient SFG fields, based on a well-known nonlinear optics formalism \cite{jankowski2024ultrafast} (a full derivation is provided in the Supplementary Material). Assuming that the gating spot is much larger than the mode profiles yields the following expression for the time-resolved SFG images:
\begin{flalign}
    \label{eq:imaging_MM_fast}
    \begin{split}
        &I_{SFG}(x,y,\tau) \propto  \int \left|G(t-\tau)\right|^2\left|\sum_{\nu}F_{\nu}(x,y)R_{\nu}(t)e^{j\omega_{\nu}  t} \right|^2dt,
    \end{split}
\end{flalign}
where $F_{\nu}(x,y)$ are the spatial profiles of the excited modes $\nu$, with slowly decaying amplitudes $R_{\nu}(t)$ and frequencies $\omega_{\nu}$. $G(t-\tau)$ is the delayed envelope of the gating pulse. Since the spectral distribution of excited cavity modes is much narrower than the pulse bandwidth, we can approximate $G(t-\tau)\sim\delta(t-\tau)$:
\begin{flalign}
    \label{eq:imaging_MM_slow}
    \begin{split}
        &I_{SFG}(x,y,\tau) \propto  \left|\sum_{\nu}F_{\nu}(x,y)R_{\nu}(\tau)e^{j\omega_{\nu}  \tau} \right|^2.
    \end{split}
\end{flalign}

The SFG spectrograms are obtained by collecting the SFG signal through the slit of a spectrometer. Assuming cavity resonances much narrower than the gate pulse bandwidth, yields:
\begin{flalign}
    \label{eq:spectrogram_MM_fast}
    \begin{split}
        &S_{SFG}(\omega,\tau) \propto  \int\left|\sum_{ \nu}F_{\nu}(0,y)R_{\nu}(\tau)e^{j\omega_{\nu}\tau}g(\omega-\omega_G-\omega_{\nu})\right|^2dy,
    \end{split}
\end{flalign}
where $g(\omega-\omega_G-\omega_\nu)$ is the gating pulse spectrum shifted to the sum-frequency domain for each cavity mode $\nu$, $\omega_G$ is the carrier frequency of the gating pulse, and the integral over $y$-coordinate accounts for full vertical binning in the imaging spectrometer. Thus, integrating the single mode spectrogram (Fig.\,\ref{fig:Figure_2}c) along the wavelength coordinate results in the decay trace of the excited mode. As seen in Fig.\,\ref{fig:Figure_2}d, the integration (black curve) fits well with an exponential decay curve (orange dashes): $A_1(\tau)=\Theta (\tau)Ae^{-\tau/\tau_0}$, with $\Theta(\tau)$ being the Heaviside function. The fitted lifetime $\tau_0$ is $156 \pm 5\,$ps (Q-factor of 62000). This translates to a $26\,\mu$eV (12\,pm) linewidth of the resonant mode, well below our spectrometer spectral resolution. Similarly, the integration of the two-modes spectrogram (Fig.\,\ref{fig:Figure_2}g) yields the temporal beating trace shown in black in Fig.\,\ref{fig:Figure_2}h. Fitting it with a two-mode beating model $A_2(\tau) =\Theta(\tau)|R_{11e}+R_{11o}e^{j(\Delta\phi+\Delta\omega\tau)}|^2e^{-\tau/\tau_{11}}$ (orange dashes) allows to reconstruct the complete dynamics of the two modes: They decay with a lifetime of $\tau_{11} = 203 \pm 0.7$\,ps (Q-factor of 80,000), a phase off-set  $\Delta\phi = 0.1 \pm 0.003\,\pi$\,rad, and an amplitude ratio  $|\frac{R_{11o}}{R_{11e}}|=0.13 \pm 0.008$.

\subsection*{Ultrafast multi-modal dynamics}\label{subsec2_2}

\begin{figure*}[ht]
\centering\includegraphics[width=\linewidth]{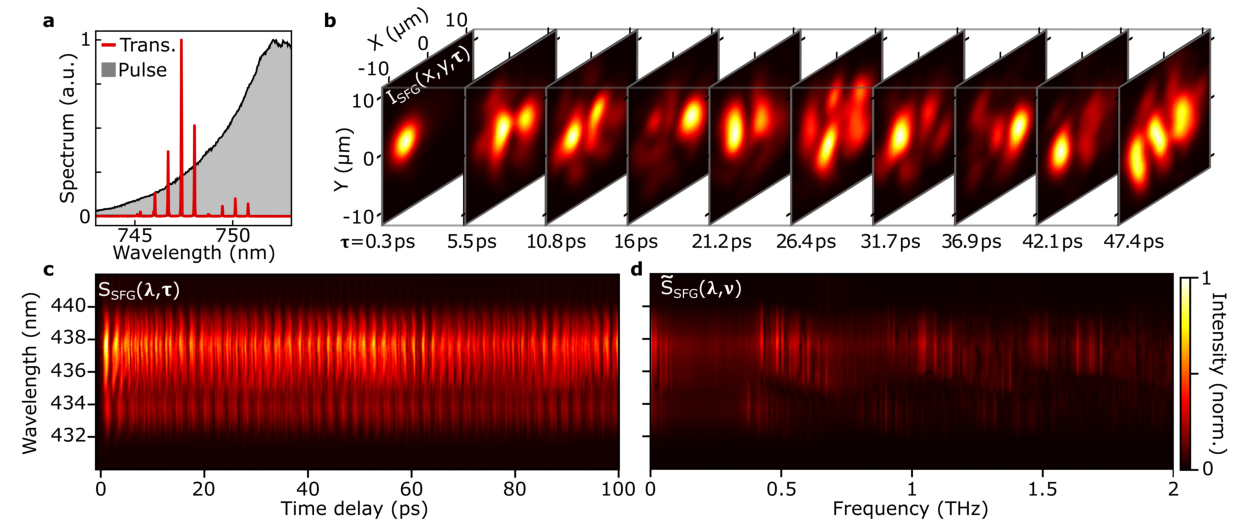}
\caption{\label{fig:Figure_3} \textbf{a.} Transmission spectrum of the cavity (red) when excited with a broad-band resonant pulse (gray). Multiple longitudinal and transverse modes are excited. \textbf{b.} Snapshots of the
transmitted SFG images at specific time-delays. \textbf{c.} SFG spectrogram.
\textbf{d.} Fourier-transform of panel (b) along the time delay coordinate. For clarity, the normalized intensities in (c) and (d) are corrected by a power of 0.5.}
\end{figure*}

Having confirmed that intra-cavity optical gating can provide both qualitative and quantitative information about the dynamics of individual resonant modes, we now explore the cavity dynamics when a full manifold of modes is excited (Fig.\,\ref{fig:Figure_3}a). The time-resolved SFG images now display complex multi-mode dynamics in the transverse field distribution, as shown in Fig.\,\ref{fig:Figure_3}b (see video in Supplementary Material). The corresponding SFG spectrogram is shown in Fig.\,\ref{fig:Figure_3}c, featuring fast oscillations along the temporal axis, with noticeable manifestations across the SFG spectrum.

This is further highlighted by taking the Fourier-transform of the spectrogram along the temporal coordinate, as shown in Fig.\,\ref{fig:Figure_3}d. Several notable features are observed: First, two spectral domains are lit-up around 434\,nm and 437\,nm, corresponding to two different excited longitudinal cavity modes, separated by 5\,THz. Second, the fundamental beating periodicity, corresponding to the heterodyning of nearest-neighbouring transverse modes with one-another, changes sensibly across the spectrum, from 0.42\,THz around 438\,nm to 0.69\,THz at 436\,nm. These changes reveal an anharmonic distribution of the transverse cavity modes, and are rooted in residual deviations of the curved micro-mirror from a perfect parabola. Third, harmonics of these frequencies are visible up to a few THz, indicating heterodyning between second-nearest (or higher order) neighbouring modes. These observations emphasizes another virtue of the spectrogram, allowing to capture signal bandwidths that are larger than the gating pulse-bandwidth thanks to time-frequency redundancy \cite{trebino_frequency-resolved_2000}. Note that a full characterization of complex intra-cavity field can be obtained by recording SFG spectrograms from each point of the $(x,y)-$plane, and applying established phase retrieval \cite{trebino_frequency-resolved_2000, eilenberger2013imaging, Shechtman2014} or machine learning algorithms \cite{zahavy2018deep}. Such data-intensive analysis, while important for the characterization of more complex optical resonators, is beyond the scope of the present work.

To further demonstrate the capabilities of this approach, we report in Extended Fig.\,\ref{fig:XFigure2_1} the time-resolved dynamics of a nominally flat monolithic TFLN microcavity (Fig.\,\ref{fig:XFigure2}), feature in-plane propagation and wavepacket dynamics. There, the sample boundaries and residual mirror curvature lead to periodic refocusing of the excited wavepacket on a picosecond time-scale. These results open new avenues for exploring complex and topological photonic band structures, where intra-cavity optical gating offers a powerful tool for probing ultrafast wavepacket dynamics.

\subsection*{Nonlinear initialization of cavity modes}\label{subsec2_3}

\begin{figure*}[!ht]
\centering\includegraphics[width=\linewidth]{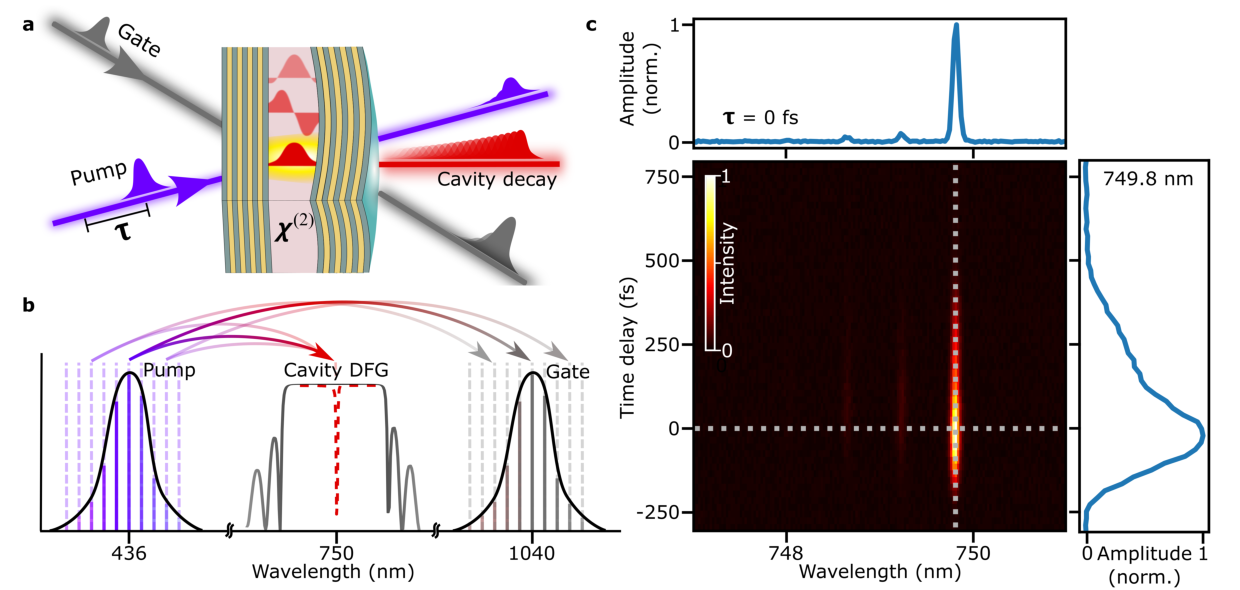}
\caption{\label{fig:Figure_4} \textbf{a.} Illustration of the stimulated intra-cavity DFG generation process. The co-propagating pump and gate pulses (angled here is for clarity) reach the cavity with a controlled delay time $\tau$. Cavity modes are populated by stimulated intra-cavity DFG when the pump and gate signals overlap in space and time within the cavity. \textbf{b.} Spectral arrangement for intra-cavity DFG. The gate pulse (1040\,nm) stimulates the DFG of resonant cavity photons near 750\,nm. \textbf{c} Time-integrated emission spectrum of the cavity, as a function of delay between the pump and gate pulses. Top and side panels show line-cuts along the gray dashed lines, corresponding to $\tau=0$\,ps and $\lambda=749.8$\,nm respectively.}
\end{figure*}

While the results presented thus far highlight optical gating for the \textit{on-demand extraction} of multimode intra-cavity dynamics, we now introduce a complementary approach that enables the \textit{on-demand injection} of optical fields into the cavity. In this scheme, depicted in Fig.\,\ref{fig:Figure_4}a, the microcavity is simultaneously excited by a pump and a gate pulse, both located outside of the DBR stop-band (150\,fs durations, 436\,nm and 1040\,nm carrier frequencies, resp.). As the two pulses traverse the microcavity, the gate stimulates difference-frequency generation (DFG) from the pump, generating photons in the resonant cavity wavelength range. Crucially, as this process occurs within the structured electromagnetic vacuum field of a 0D, high finesse microcavity, the excess energy can only be carried away through the generation of photons at the discrete cavity mode frequencies (Fig.\,\ref{fig:Figure_4}b).

The measured spectrum, emitted by the cavity around its resonant wavelengths, is shown in Fig.\,\ref{fig:Figure_4}c, as a function of the time-delay between the pump and gate pulses. The attached cross-sectional insets clearly demonstrate the instantaneous injection of a series of transverse cavity modes ($\sim0.6$\,nm mode spacing), when both the pump and gate pulses are simultaneously present in the microcavity. The span of modes excited through this process is determined by the spectral overlap between the cavity mode spectrum and the convolved pump and gate spectra, as well as by modal overlap between the cavity modes and the pulses. Spectral selectivity can thus be improved by narrowing the bandwidths of the pump and gate pulses, at the expense of an increased uncertainty over the instantiation time of the cavity modes. Alternatively, temporal and spatial pulse shaping may be employed to improve overlap with certain target intra-cavity state~\cite{heuck2020controlled}. This technique provides an efficient pathway for all-optical, ultrafast control of cavity excitation dynamics, fully compatible with ultra-high-finesse DBR resonators.

\section*{Discussion}

We have demonstrated a novel platform for ultrafast nonlinear optics in high-finesse Fabry-Perot microcavities, enabled by the integration of TFLN with a tunable 0D DBR resonator. This architecture combines long-lived optical resonances with femtosecond dynamical control via instantaneous second-order nonlinear processes. Using this platform, we achieve ultrafast optical gating --both injection and extraction-- of cavity fields, offering coherent access to complex intra-cavity dynamics.

A key advantage of this approach is its ability to extract and instantiate the intra-cavity field with control over its spatiotemporal profile, enabling selective excitation and readout of optical modes. This makes the technique broadly applicable to more complex photonic structures, such as coupled cavity arrays~\cite{st2017lasing, klembt2018exciton}, with direct relevance to polaritonics~\cite{sanvitto2016road}, photon Bose–Einstein condensates (BEC)~\cite{klaers2010bose}, and optical thermodynamics~\cite{dinani2025universal}.

Importantly, the open architecture of our nonlinear microcavity allows for straightforward integration of diverse material systems, including epitaxial thin films via wafer bonding, 2D materials through van der Waals assembly, and molecular dyes via spin-coating. This flexibility paves the way toward hybrid $\chi^{(2)}$-$\chi^{(3)}$-nonlinear photonic systems. We further strengthen this point by demonstrating, in Extended Fig.\,\ref{fig:XFigure3} and Fig.\,\ref{fig:XFigure4}, ultrafast optical gating in a high finesse tunable microcavity, assembled within a close-loop cryostat operating at 4\,K. 

Another important aspect of the demonstrated technique is the efficiency of the nonlinear interactions. While this work focuses on ultrafast, multimode cavity dynamics using broad femtosecond gate pulses, higher conversion efficiencies can be achieved through spatial and temporal pulse shaping to mode-match specific intra-cavity fields~\cite{heuck2020controlled}, as discussed in the Supplementary Material. Efficient depletion of cavity modes using tailored gate pulses could offer precise control over the non-Hermitian evolution of intra-cavity states, enabling studies of position- and time-dependent quenches in driven-dissipative photonic systems.

Finally, in the limit of deterministic quantum state extraction, intra-cavity optical gating could enable manipulation and study of quantum optical states in a well-defined temporal mode basis --an essential requirement for various all-optical quantum information protocols. Notably, the second-order correlation function $g^{(2)}(\tau)$ of the intra-cavity field is faithfully up-converted through the SFG process, even at finite conversion efficiencies, and should be directly measurable with the current device architecture. This platform thus constitutes a powerful tool for the advancing ultrafast quantum optics in integrated photonics systems.

\section*{Methods}\label{secM}

\subsection*{Device fabrication}\label{subsecM_1}
The main device presented in the text was fabricated in two parts. The DBR mirror (quarter-wave stack of Ta$_2$O$_5$ and SiO$_2$) was sputter-coated onto a silica wafer by FiveNine Optics Inc., USA. A slab of LiNbO$_3$ (NGK, Japan) was then wafer-bonded onto the DBR, and then polished to a thickness of 10 $\mu m$. The bonded wafer was then diced into 7\,mm x 7\,mm chips. The curved DBR was sputtered onto an ablated SiO$_2$ mesa chip (Qlibri GmbH, Germany), realizing a curved DBR with radius of curvature of about 30 $\mu m$. The two DBRs where brought into proximity using micro-manipulators and piezo-actuators. 

The device used in the cryogenic setup was based on a similar flat TFLN layer bonded to a DBR chip. For this experiment, the curved mirror was sputtered onto the ablated facet of a single mode optical fiber (Qlibri GmbH, Germany).

The monolithic cavity device, shown in Supplementary Data, was fabricated from a 7 $\mu m$ thick TFLN flake, chipped off a commercially available LiNbO$_3$ wafer (NanoLN Ltd., China), which was attached to a flat DBR (FiveNine Optics Inc., USA) using a 150\,nm thick layer of PMMA. Subsequent sputtering of the top DBR (FiveNine Optics Inc., USA) completed the cavity. Stress transfer during the top mirror deposition resulted in slight buckling of the flak, yielding a nominally flat Fabry-Perot microcavity, with a mirror radius of curvature of the order of millimeters.

\subsection*{Time-integrated spectroscopy and imaging}\label{subsecM_2}
For the time-integrated spectra and images we illuminate the cavity with the broadband pulses (750\,nm, 10\,nm FWHM bandwidth, 500\,mW average power impinging at the input of the cavity) sourced from the tunable output of a Chameleon Discovery NX laser (Coherent Inc., USA). Most of the power is reflected back by the DBRs, and only light resonant with the cavity is transmitted. The transmission spectrum is collected with an imaging spectrometer (SpectraPro HRS-750, Princeton Instruments Inc., USA) equipped with a charge-coupled device (CCD) camera (ProEM-HS, Princeton Instruments Inc., USA). To resolve the spectrum and the spatial distributions of the modes in the same measurement, we use the horizontal coordinate of the CCD for the spectrum, the vertical coordinate of the CCD for the spatial $y$-coordinate, and sweep the horizontal position of the tube lens at the spectrometer input to scan the $x$-coordinate. Thus, we acquire a three-dimensional data set.
The spatial coordinates were calibrated against microscopically measured dimensions.

The transmission spectra recorded in the cryogenic setup utilized a superluminescent light-emitting diode (Exalos AG., Switzerland) as a broadband spectroscopic light source.

\subsection*{Time-resolved spectroscopy and imaging}\label{subsecM_3}
To generate the time-resolved data we tune the delay between two pulses generated by the same pulsed laser (Chameleon Discovery NX). Its tunable output is used to seed the cavity with the resonant modes (pump pulse) and the fixed wavelength output at 1040\,nm is used as the gating pulse. To produce the clear signals shown in the text we used $\sim500$\,mW average power over the full bandwidth of the tunable pump, whereas the gating pulse fluence was 0.2\,mJ/cm$^2$. Both pulses were polarized along the extra-ordinary axis of the LiNbO$_3$ crystal.
For the stimulated loading of the cavity by DFG, we doubled the tunable output of the laser (tuned to 872\,nm) using a Beta-Barium Borate (BBO) crystal to generate a 436\,nm pulse. The pump signal average power was 150\,mW, while the gating fluence was enhanced by focusing to 0.6\,mJ/cm$^2$. Both pulses were polarized along the extra-ordinary axis of the LiNbO$_3$ crystal.
Collecting the SFG light is done in two modalities:
\textit{Imaging}: The spectrometer slit is wide open, the grating is tuned to zeroth-order scattering, and the CCD array is used as a camera.
\textit{spectroscopy}: The spectrometer slit is closed, and the CCD is fully vertically binned to generate a spectrum at each time delay between the pump and the gating pulses.

\subsection*{Cryogenic setup}\label{subsecM_4}
To work at cryogenic temperatures, the cavity setup was built into a cryostat (AttoDRY800, Attocube GmbH, Germany) equipped with free-space and fiber optical ports. The flat DBR mirror of the cavity was held static, while the fiber DBR was actuated. To control this motion, we used a cryo-compatible piezo-electric stage (CPSHR1, JPE, Netherlands, see Ref. \cite{childress2021}) that also passively damped the vibrations in the setup. The pump pulse average power was set to 13\,mW, while the gate pulse fluence was reduced to 20\,$\mu J/cm^2$. 

\newpage

\section*{Extended Figure 1: Gating pulse spectrum}\label{X_fig1}

\begin{figure*}[!ht]
\centering\includegraphics[width=7cm]{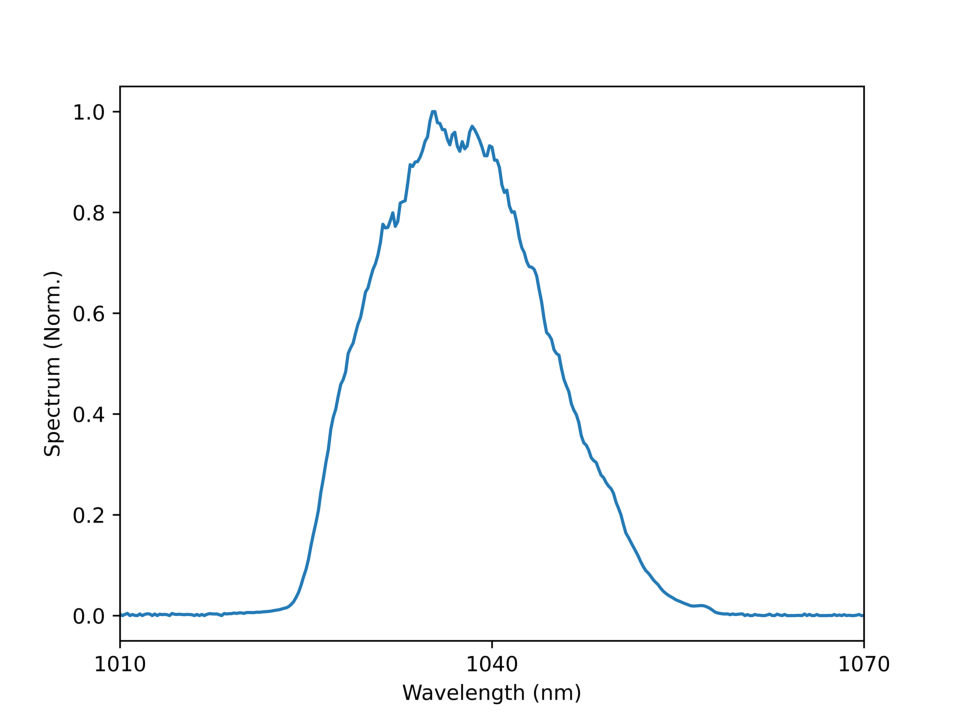}
\caption{\label{fig:XFigure1} Spectrum of the gating pulse used in the experiments, integrated using an optical spectrum analyzer (Yokagawa3670).
}
\end{figure*}

\newpage

\section*{Extended Figure 2: Monolithic microcavity image}\label{X_fig2}

\begin{figure*}[!ht]
\centering\includegraphics[width=7cm]{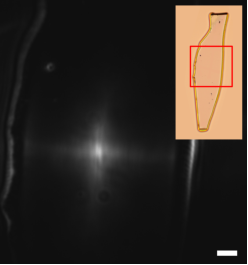}
\caption{\label{fig:XFigure2} Micrograph of the nominally flat monolithic microcavity. The edges of the 7\,$\mu$m thick TFLN flake define the transverse shape of the cavity together with the residual curvature of the flake. The illumination source is a 150\,fs pulse centered at 750\,nm. The central spot is defined by its overlap with the resonant cavity modes. Light scattering off the edges of the flake is visible on the left and the right edges of the image. The scale bar is 10\,$\mu$m. Inset: a microscope image of the flake. The red rectangle marks the area imaged in the main panel.  
}
\end{figure*}

\newpage

\section*{Extended Figure 3: Monolithic cavity time-resolved characterization}\label{X_fig3}

\begin{figure*}[!ht]
\centering\includegraphics[width=\linewidth]{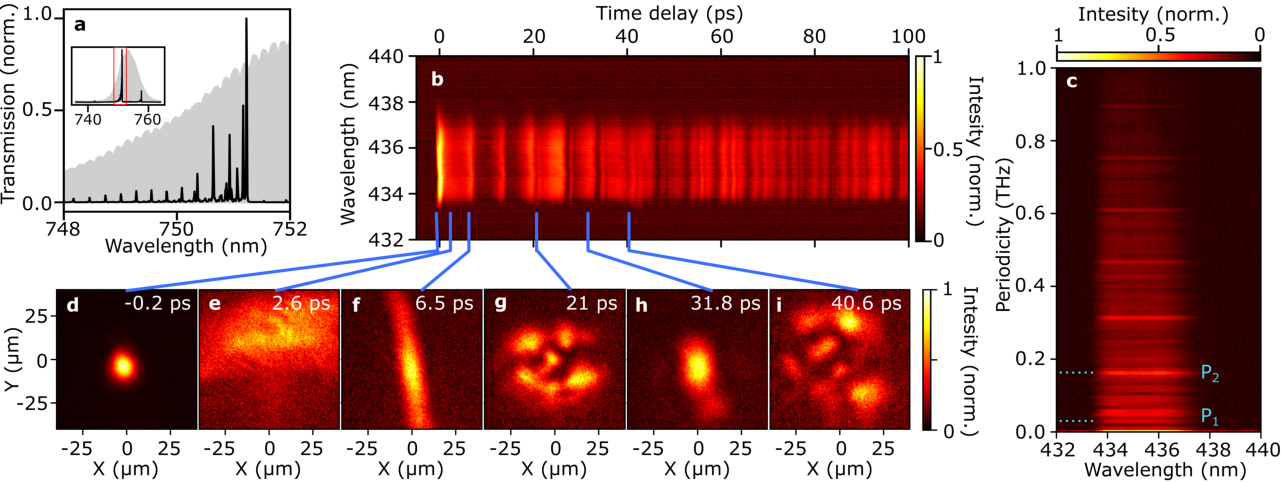}
\caption{\label{fig:XFigure2_1} Time-resolved characterization of multi-mode dynamics in the nominally flat monolithic microcavity. \textbf{a.}  Transmission spectrum of the cavity (black) when excited with an un-filtered resonant pulse (gray shade). Multiple lines are excited. The inset shows a zoomed-out spectrum, showing additional longitudinal mode manifolds in the cavity spectrum. \textbf{b.} SFG spectrogram, showing periodic beating across the SFG spectrum. For clarity, the normalized intensity is raised by a power of 0.5. \textbf{c.} Fourier-transform of panel (b) with respect to the temporal coordinate. The beating periodicity is uniform across the spectrum. Two fundamental frequencies are highlighted. $P_1 = 0.04\,$THz related to light echoing in the cavity along its long north-south axis. $P_2 = 0.17\,$THz relates to light echoing along the short east-west axis. \textbf{d-i.} Snap-shot images of the SFG signal at specific time-delays, showing the wavepacket dynamics across the cavity (see full movie in Supplementary Material).
}
\end{figure*}

\newpage

\section*{Extended Figure 4: Cryogenic cavity setup - Time integrated characterization}\label{X_fig4}

\begin{figure*}[!ht]
\centering\includegraphics[width=\linewidth]{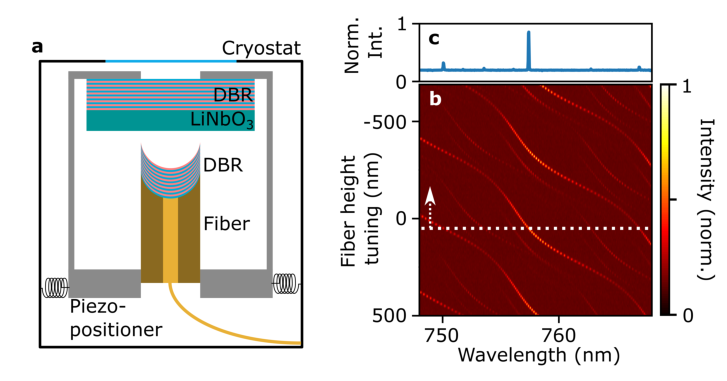}
\caption{\label{fig:XFigure3} \textbf{a.} Schematic of the cryogenic cavity setup. The curved DBR is coated on top of an ablated optical fiber, held at a controlled position away from the flat DBR carrying the slab of TFLN, by using a vibration-isolated piezo-positioners \cite{childress2021}. The entire assembly resides inside a 4\,K closed-loop cryostat. Light is coupled in through the top window of the cryostat and through the flat DBR. The transmitted light is collected through the curved DBR and the fiber attached to it.  \textbf{b.} Time-integrated transmission spectrum of the cavity, at 4\,K temperature, upon scanning of the fiber distance away from the flat DBR. The most visible spectral lines are associated with different longitudinal mode orders in the cavity, with free spectral range of about 7.5\,nm, while the fainter lines correspond to higher transverse modes of the 0D cavity. The curved dispersions and variations in modes brightness are due to hybridization between air-like and TFLN-like cavity modes. \textbf{c.} Spectral cross-section along the dashed line marked in (b).  
}
\end{figure*}

\newpage

\section*{Extended Figure 5: Cryogenic cavity setup - Time resolved characterization}\label{X_fig5}

\begin{figure*}[!ht]
\centering\includegraphics[width=\linewidth]{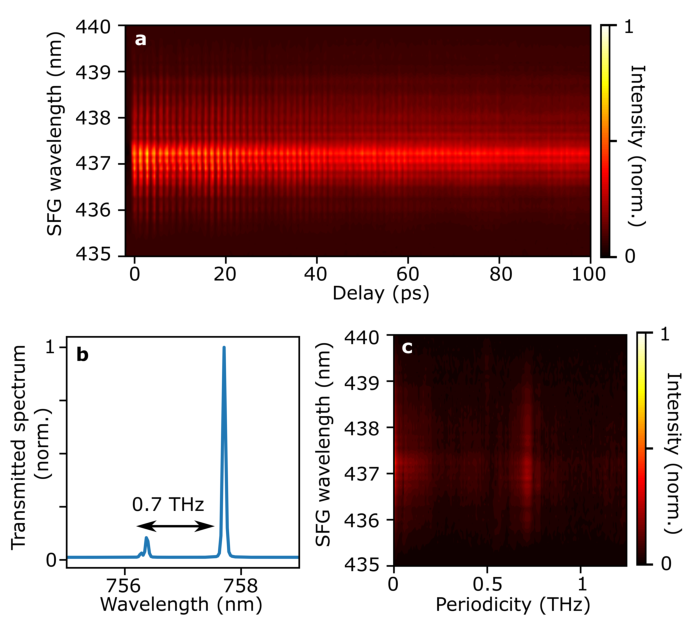}
\caption{\label{fig:XFigure4} \textbf{a.} Time-resolved SFG spectrum acquired inside the cryostat at 4\,K temperature. Fast beatings are observed.  \textbf{b.} Time-integrated transmission spectrum of the pump pulse. Only three excited transverse modes are collected by the fiber. The spectral splitting between the fundamental mode and the weak pair of higher-order modes is about 0.7\,THz, in accord with the beating frequency observed in (a).  \textbf{c.} A Fourier transform of the spectrogram in (a). The main beating frequency appears near 0.7\,THz.}
\end{figure*}

\newpage

\section*{Supplementary derivations}

\subsection*{Intra-cavity SFG derivation}

We start by describing the dependence of the SFG on the gating pulse and the resonant intra-cavity field as a basis for the analysis of our experimental results. The derivation relies on similar procedures described by Jankowski et al. in Ref. \cite{jankowski2024ultrafast}. It is based on the following assumptions. The diffraction in the cavity is small, so the transverse profile of the fields does not change along the cavity. As well, the transient envelopes are assumed to be slow compared with the carrier waves, and finally the gating field and the resonant modes are assumed to be un-depleted in the SFG process. 
The fields are then defined along the following ansatz:

\begin{align}
    \label{eq:fields}
    &E_{gate}(t) = \frac{1}2G(t-\tau)F_G(x,y)e^{j(\omega_G(t-\tau)-k_Gz)} + c.c.  \\
    % &E_{SFG} = \frac{1}2 S(z,t)F_S(x,y)e^{j(\omega_st-k_sz)} + c.c.  
    &E_{res}(t) = \frac{1}4 \sum_{m,\mu}R_{m,\mu}(t)F_{\mu}(x,y)[e^{j(\omega_{m,\mu}t-k_R^{m,\mu}z)}+e^{j(\omega_{m,\mu}t+k_R^{m,\mu}z)}] + c.c. \\
    &E_{SFG}(t) = \frac{1}2\sum_{\nu}S_{\nu}(z,t)F_{\nu}(x,y)e^{j(\omega_st-k_s^\nu z)} + c.c.  ,
\end{align}
and in frequency domain:
\begin{align}
    \label{eq:fields_freq}
    &E_{gate}(\omega) = g(\omega-\omega_G)F_G(x,y)e^{-j((\omega-\omega_G)\tau+k_Gz)}  \\
    % &E_{SFG} = \frac{1}2 S(z,t)F_S(x,y)e^{j(\omega_st-k_sz)} + c.c.  
    &E_{res}(\omega) = \frac{1}2 \sum_{m,\mu}R_{m,\mu}\delta(\omega-\omega_{m,\mu})F_{\mu}(x,y)[e^{-jk_R^{m,\mu}z}+e^{jk_R^{m,\mu}z}] \\
    &E_{SFG}(\omega) = \sum_{\nu}S_{\nu}(z,\omega-\omega_s)F_{\nu}(x,y)e^{-jk_s^\nu z} . 
\end{align}

We assign the gating pulse with a single forward-propagating mode at $\omega_G$ with wave-vector $k_G$, a wide Gaussian beam profile $F_G(x,y)$, and a delayed temporal envelope $G(t-\tau)$ that does not change along the cavity. The resonant field is defined as a superposition of cavity eigenmodes, of longitudinal order $m$ and transverse order $\mu$. Their transverse mode profiles are $F_{\mu}(x,y)$. Their standing-wave profiles are set with frequencies $\omega_{m,\mu}$, and wave-numbers $k_R^{m,\mu}$. The modal amplitudes $R$ are very slow functions of time, decaying over much longer times than the gating pulse width, justifying their description as delta-functions in the spectrum. We will use this approximation hereafter. Their occupation is determined by some excitation pulse at $t=0$. Finally, we span the SFG signal using the same transverse profiles as the cavity resonant modes, but let them propagate uni-directionally at a frequency $\omega_s$ with wave-vector $k_s^\nu$. Their temporal envelopes $S_\nu(z,t)$ are allowed to evolve as the waves propagate in the cavity.

Since the SFG process requires phase-matching, we write the propagation equation of the SFG modes using only forward propagating waves, as follows:

\begin{align}
    \label{eq:propagation}
    \partial_zS_{\nu}(z,\omega-\omega_s) = -\frac{j\omega_s}P e^{jk_s^{\nu}z}\iint F_{\nu}^*(x,y)P_{nl}dA,
\end{align}
\begin{align}
    \label{eq:P_nl}
    \begin{split}
        &P_{nl} = \frac{\epsilon_0\chi^{(2)}_{zzz}}{2\pi}\int_{-\infty}^{\infty}
        E_{res}^{forward}(\omega_2)E_{gate}(\omega-\omega_2)d\omega_2 = \\
        &=\frac{\epsilon_0\chi^{(2)}_{zzz}}{4\pi} \sum_{m,\mu}R_{m,\mu}g(\omega-\omega_G-\omega_{m,\mu})e^{-j(\omega-\omega_G-\omega_{m,\mu})\tau}e^{-j(k_R^{m,\mu}+k_G)z}F_{\mu}(x,y)F_G(x,y) , 
    \end{split}
\end{align}
where P is the power unit carried by the SFG mode profile for amplitude of $S(z,t) = 1$. Substituting \ref{eq:P_nl} in \ref{eq:propagation}, there appears a spatial overlap integral $\iint F_{\nu}^*F_{\mu}F_G dA$. Since it is assumed that the gating beam is much wider than the resonant modes, this integral becomes an overlap integral between mode profiles $F_{\nu}$ and $F_{\mu}$, zeroing for all but for $\mu = \nu$, where the result is the normalization power $P$ times the wave impedance $Z_s$. 
Thus, the evolution equation of the SFG amplitude can be simplified to:
\begin{flalign}
    \label{eq:propagation_simple}
    \begin{split}
        &\partial_zS_{\nu}(z,\omega-\omega_s) = \\
        &=-\frac{j\epsilon_0 \omega_s \chi^{(2)}_{zzz}Z_s}{\pi}  \sum_{m}R_{m,\nu}F_G(0,0)g(\omega-\omega_s-\Delta\omega_{m,\nu})e^{-j(\omega-\omega_s-\Delta\omega_{m,\nu}) \tau}e^{-j\Delta k_{m,\nu}z},
    \end{split}
\end{flalign}
Where we define the momentum-mismatch term as $\Delta k_{m,\nu} = k_R^{m,\nu} + k_G - k_s^{\nu}$, and the detuning term $\Delta\omega = \omega_{m,\nu}+\omega_G-\omega_s$. 

Now we can express the momentum-mismatch as function of $\omega$. Expanding it as a Taylor series reveals the dominant terms: the momentum-mismatch of the carrier waves and the pulse walk-off:
\begin{flalign}
    \label{eq:k_taylor}
    \Delta k = \Delta k_0 + (\omega-\omega_s) \Delta\partial_\omega k.
\end{flalign}

Evaluating these terms for the experimental system at hand ($L = 10\,\mu$m-thick LiNbO$_3$ flake, and 436\,nm, 750\,nm, and 1040\,nm for the SFG, resonance, and gating field wavelengths, respectively) gives:
\begin{flalign}
    \label{eq:k_taylor_values}
    &\Delta k_0 L= 5.4 \pi ; \ \ \ \ \Delta\partial_\omega k L = 12 fs.
\end{flalign}

Thus, while the carrier phase mismatch is significant, the pulse walk-off is negligible compared to the original $\sim150$\,fs pulse widths. This allows to simplify the rest of the derivation to get the SFG mode at the output facet of the cavity ($z = L$):
\begin{flalign}
    \label{eq:final_Omega}
    \begin{split}
    s_\nu(L,\Omega) &= -\frac{j\epsilon_0 \omega_s \chi^{(2)}_{zzz}Z_sF_G(0,0)}{\pi} \times\\ &\times\sum_{m}R_{m,\nu}g(\omega-\omega_s-\Delta\omega_{m,\nu})e^{-j(\omega-\omega_s-\Delta\omega_{m,\nu}) \tau}\int_{0}^{L} e^{-j\Delta k_{0m,\nu}z} dz.
     \end{split}
\end{flalign}

Defining the effective phase-matched interaction length as $\mathcal{L}_{m,\nu} = \int_{0}^{L} e^{-j\Delta k_{0m,\nu}z} dz$ and transforming the equation to the time-domain gives:
\begin{flalign}
    \label{eq:final_t}
    \begin{split}
        S_\nu(L,t) &= -{j\epsilon_0 \omega_s \chi^{(2)}_{zzz}Z_sF_G(0,0)e^{-j\omega_G \tau}}  \sum_{m}R_{m,\nu}(t)G(t-\tau)e^{j\Delta\omega_{m,\nu} t}\mathcal{L}_{m,\nu}.
    \end{split}
\end{flalign}

The SFG field at the output facet of the cavity is then expressed by:
\begin{flalign}
    \label{eq:Esfg}
    \begin{split}
        E_{SFG}(x,y,t,\tau) &= {-j\epsilon_0 \chi^{(2)}_{zzz}\omega_sZ_s}F_G(0,0)G(t-\tau)\times \\
        &\times\sum_{m, \nu}F_{\nu}(x,y)R_{m,\nu}(t)e^{j\Delta \omega_{m,\nu} t}e^{j(\omega_s  t -k_s^{\nu}L)}\mathcal{L}_{m,\nu},
    \end{split}
\end{flalign}
and after some rearranging: 
\begin{flalign}
    \label{eq:final_Esfg}
    \begin{split}
        E_{SFG}(x,y,t,\tau) &= {-j\epsilon_0 \chi^{(2)}_{zzz}\omega_sZ_s}F_G(0,0)G(t-\tau)e^{j\omega_G t}\times \\
        &\times\sum_{m, \nu}F_{\nu}(x,y)R_{m,\nu}(t)e^{j\omega_{m,\nu} t}e^{-jk_s^{\nu}L}\mathcal{L}_{m,\nu},
    \end{split}
\end{flalign}
with  $k_s = \frac{\omega_G+\omega_R}c$, $\omega_R$ being some representative frequency in the resonant domain.

This expression states that the SFG signal is a product of the gating pulse with a coherent superposition of cavity modes, weighted by their occupation amplitudes and individual effective interaction lengths. Rather than oscillating at frequency $\omega_s$, these modes oscillate at $\omega_{m,\nu}+\omega_G$, as expected from the SFG process. 

\subsection*{SFG imaging - theoretical derivation}

The SFG images are expressed as an accumulation over time of the SFG intensity. It is assumed here for simplicity that the spatial coordinates are unchanged by the relay of the image from the cavity to the detector (for simplicity, we use the index $\nu$ for all the modes of the cavity, longitudinal and transverse):
\begin{flalign}
    \label{eq:imaging}
    \begin{split}
        &I_{SFG}(x,y,\tau) = \int\left|E_{SFG}(x,y,t,\tau)\right|^2dt = \\
        &=\left( {\epsilon_0 \chi^{(2)}_{zzz}\omega_sZ_sF_G(0,0)}\right)^2  \int \left|G(t-\tau)\right|^2\left|\sum_{\nu}F_{\nu}(x,y)R_{\nu}(t)e^{j\omega_{\nu}  t}e^{-jk_s^{\nu}L}\mathcal{L}_{\nu} \right|^2dt.
    \end{split}
\end{flalign}
Assuming the effective interaction length and accumulated phases are similar across the examined mode spectrum ($e^{-jk_s^\nu L} \approx e^{-jk_sL} $, and  $\mathcal{L}_{\nu} \approx \mathcal{L}_R$) gives the expression in Eq. \ref{eq:imaging_MM_fast} in the main text:
\begin{flalign}
    \label{eq:imaging_simple}
    \begin{split}
        &I_{SFG}(x,y,\tau) 
        =\left( {\epsilon_0 \chi^{(2)}_{zzz}\omega_sZ_sF_G(0,0)|\mathcal{L}_{R}|}\right)^2  \int \left|G(t-\tau)\right|^2\left|\sum_{\nu}F_{\nu}(x,y)R_{\nu}(t)e^{j\omega_{\nu}  t} \right|^2dt.
    \end{split}
\end{flalign}
This is a convolution of the gating pulse envelope with the transient image of the cavity modes. For example, a very long gating pulse, with narrow bandwidth, would create an average over all the oscillations, loosing the temporal resolution. A broadband pulse, on the other hand, would sample the mode heterodyning at a specific time delay $\tau$ yielding Eq. \ref{eq:imaging_MM_slow} in the main text:
\begin{flalign}
    \label{eq:imaging_delta_final}
    \begin{split}
        &I_{SFG}(x,y,\tau) =\left( {\epsilon_0 \chi^{(2)}_{zzz}\omega_sZ_sF_G(0,0)G(0)|\mathcal{L}_{R}|}\right)^2 \left|\sum_{ \nu}F_{\nu}(x,y)R_{\nu}(\tau)e^{j\omega_{\nu}  \tau} \right|^2.
    \end{split}
\end{flalign}

\subsection*{SFG Spectroscopy - theoretical derivation}

The SFG spectrum can inform about the cavity mode spectrum and its transient behavior, rather than on the mode spatial profiles. It is taken by dispersing the spectrum of the SFG image from a specific local environment (for simplicity, we use the index $\nu$ for all the modes of the cavity, longitudinal and transverse):
\begin{flalign}
    \label{eq:spectrum}
    \begin{split}
        &S_{SFG}(\omega,\tau, x, y) = \left|\int E_{SFG}(x, y,t,\tau)e^{-j\omega t}dt\right|^2 = \\
        &=\left( {\epsilon_0 \chi^{(2)}_{zzz}\omega_sZ_sF_G(0,0)}\right)^2 \times\\
        &\times\left|\sum_{ \nu}R_{\nu}(\tau)F_{\nu}(x,y)e^{-j(\omega - \omega_G-\omega_{\nu} )\tau}g(\omega-\omega_G-\omega_{\nu})e^{-jk_s^{\nu}L}\mathcal{L}_{\nu}\right|^2 = \\
        &=\left( {\epsilon_0 \chi^{(2)}_{zzz}\omega_sZ_sF_G(0,0)}\right)^2 \left|\sum_{ \nu}R_{\nu}(\tau)F_{\nu}(x,y)e^{j\omega_{\nu}\tau}g(\omega-\omega_G-\omega_{\nu})e^{-jk_s^{\nu}L}\mathcal{L}_{\nu}\right|^2,
    \end{split}
\end{flalign}
which is a coherent sum of replica of the gating spectrum $g(\omega-\omega_G)$, shifted by $\omega_{\nu}$. Another way to express this is a convolution of $g(\omega-\omega_G)$ with the modal spectrum, where each mode is assigned a delta-function in frequency: $\sum_{\nu}F_{\nu}(x,y)R_{\nu}\delta(\omega-\omega_{\nu})e^{j\omega_{\nu}\tau}$.

Specifically, we measure the spectrum using an imaging spectrometer through a narrow vertical slit (along the \textit{y}-axis, at $x=0$), and fully bin the reading:
\begin{flalign}
    \label{eq:spectrum_binned}
    \begin{split}
        &S_{SFG}(\omega,\tau)= \\ 
        &=\left( {\epsilon_0 \chi^{(2)}_{zzz}\omega_sZ_sF_G(0,0)}\right)^2 \int \left|\sum_{\nu}R_{\nu}(\tau)F_{\nu}(0,y)e^{j\omega_{\nu}\tau}g(\omega-\omega_G-\omega_{\nu})e^{-jk_s^{\nu}L}\mathcal{L}_{\nu}\right|^2dy.
    \end{split}
\end{flalign}
Here too approximating uniform effective interaction length and phase factor gives Eq. \ref{eq:spectrogram_MM_fast} from the main text:
\begin{flalign}
    \label{eq:spectrum_simple_fast}
    \begin{split}
        &S_{SFG}(\omega,\tau) =\\
        &=\left( {\epsilon_0 \chi^{(2)}_{zzz}\omega_sZ_sF_G(0,0)|\mathcal{L}_{R}|}\right)^2 \int\left|\sum_{ \nu}R_{\nu}(\tau)F_{\nu}(0,y)e^{j\omega_{\nu}\tau}g(\omega-\omega_G-\omega_{\nu})\right|^2dy.
    \end{split}
\end{flalign}
Once again, the frequency dependence in Eq. \ref{eq:spectrum_simple_fast} can be simplified if we are interested in field dynamics that are much longer-lasting than the gating pulse itself. In that case, the gate spectrum can be pulled out of the summation:
\begin{flalign}
    \label{eq:spectrum_simplified_slow}
    \begin{split}
        &S_{SFG}(\omega,\tau) =  \\
        &=\left( {\epsilon_0 \chi^{(2)}_{zzz}\omega_sZ_s|\mathcal{L}_R|F_G(0,0)}\right)^2 \left|g(\omega-\omega_G-\omega_R)\right|^2 \int\left|\sum_{ \nu}F_{\nu}(0,y)R_{\nu}(\tau)e^{j\omega_{\nu}\tau}\right|^2dy.
    \end{split}
\end{flalign}
The expression \ref{eq:spectrum_simplified_slow} is now proportional to the \textit{instantaneous} resonant field intensity.  For example, if operating in a multi-mode fashion, this will express heterodyning of the cavity modes.

\subsection*{Calculation of frequency conversion efficiency}
Assessing the SFG efficiency is done here by expressing the output SFG energy extracted from a single mode in the cavity, seeded with a single photon at $t = 0$, using an arbitrarily powered gating pulse. The SFG energy in the cavity is:
\begin{flalign}
    \label{eq:output_energy}
    \begin{split}
        &\mathcal{E}_{SFG} = \int\left[\frac{1}{2Z_s}\iint\left|E_{SFG}(x,y,t,\tau)\right|^2dxdy\right]dt = \\
        &=\frac{\left( {\epsilon_0 \chi^{(2)}_{zzz}\omega_sZ_s|\mathcal{L}_{0}|F_G(0,0)}\right)^2}{2Z_s}\int\left(G(t-\tau)R_0(t)\right)^2dt  \iint \left|F_{0}(x,y) \right|^2dxdy.
    \end{split}
\end{flalign}
The transverse spatial integral can be evaluated in relation to the single-photon occupation of the mode:
\begin{flalign}
    \label{eq:mode_single_photon}
    \begin{split}
        &\mathcal{E}_{R} = \frac{1}{2Z_0}\int R_0(t)^2dt  \iint \left|F_{0}(x,y) \right|^2dxdy = \hbar\omega_0,
    \end{split}
\end{flalign}
resulting in:
\begin{flalign}
    \label{eq:output_energy2}
    \begin{split}
        &\mathcal{E}_{SFG} = \left( {\epsilon_0 \chi^{(2)}_{zzz}\omega_s|\mathcal{L}_{0}|F_G(0,0)}\right)^2Z_sZ_0\hbar\omega_0\frac{\int\left(G(t-\tau)R_0(t)\right)^2dt}{\int R_0(t)^2dt} . 
    \end{split}
\end{flalign}
This highlights the idea that the gating pulse shape $G(t)$ can be optimized for conversion efficiency by maximizing the ratio $\frac{\int\left(G(t-\tau)R_0(t)\right)^2dt}{\int R_0(t)^2dt}$ for a given gating pulse power and a desired pulse shape of the SFG signal \cite{heuck2020photon}.
In the current case, the gating pulse is much shorter than the single mode lifetime in the cavity, and therefore this ratio can be approximated as (also assuming an exponential decay of the cavity mode):
\begin{flalign}
    \label{eq:time_int_ratio}
    \begin{split}
        &\frac{\int\left(G(t-\tau)R_0(t)\right)^2dt}{\int R_0(t)^2dt} = \frac{e^{-\tau/\tau_0}}{\tau_0}\int G(t)^2dt.
    \end{split}
\end{flalign}

Then, to calculate the efficiency of up-conversion, the SFG pulse energy is divided by the gating pulse energy:
\begin{flalign}
    \label{eq:Gating_energy}
        &\mathcal{E}_{G} = \frac{1}{2Z_G}\iint F_G(x,y)^2dxdy\int G(t)^2dt,  
\end{flalign}
yielding:
\begin{flalign}
    \label{eq:efficiency1}
    \begin{split}
        &\eta_{SFG} = 2\left( {\epsilon_0 \chi^{(2)}_{zzz}\omega_s|\mathcal{L}_{0}|}\right)^2Z_sZ_0Z_G\hbar\omega_0\frac{e^{-\tau/\tau_0}}{\tau_0}\frac{F_G(0,0)}{\iint F_G(x,y)^2dxdy},
    \end{split}
\end{flalign}
where the last ratio in the expression above is the inverse of the gating pulse spot area. Assuming a Gaussian beam profile of width $w_G$ gives:
\begin{flalign}
    \label{eq:efficiency_final}
    \begin{split}
        &\eta_{SFG} = 2\left( {\epsilon_0 \chi^{(2)}_{zzz}\omega_s|\mathcal{L}_{0}|}\right)^2Z_sZ_0Z_G\hbar\omega_0\frac{e^{-\tau/\tau_0}}{\tau_0}\frac{1}{\pi\textit{w}_G^2},
    \end{split}
\end{flalign}
with the maximal efficiency obtained for $\tau = 0$.

The table below summarizes the parameters in use to describe the current experimental details. These yield an efficiency of \textit{c.a.} $1\times10^{-16}$, corresponding to the probability of up-converting a single cavity photon using a single gate photon. In other words, it takes \textit{c.a.} $1\times10^{16}$ gating photons (about 1\,mJ gating pulse energy), to ensure a single SFG photon. Nevertheless, the conversion efficiency can be greatly improved, as the current experiment was not optimized for it, but rather for instantaneous and local characterization of the cavity modes. First, focusing the gating beam and matching it with the resonating mode profile, will raise the efficiency in proportion to the gating beam area. Second, as seen in Eq.\,S\ref{eq:output_energy2}, the gating pulse temporal shape can be better matched to deliver higher efficiencies, potentially gaining up to two orders of magnitude in efficiency. The phase-matching condition in the cavity can also be improved, by using different thickness of TFLN, or by choosing a different set of wavelengths whose coherence length is comparable or larger than the current TFLN thickness. Finally, this architecture can be extended to other $\chi^{(2)}$ materials with potentially higher nonlinear coefficients.

\begin{table}[!ht]
\caption{\label{table:S1} Experimental parameters used to calculate the frequency conversion efficiency.
}
\centering
    \begin{tabular}{|l|c|c|}
        \hline
        Parameter & Value \\
        \hline
        $\omega_0$ ($10^{12}$ rad/sec) & 2513\\
        $\omega_G$ ($10^{12}$ rad/sec) & 1812\\
        $\omega_s$ ($10^{12}$ rad/sec) & 4325\\
        $\mathcal{L}_{0}$ ($\mu m$) & 0.95\\
        $\textit{w}_G$ ($\mu m$) & 15 \\
        $Z_s$ ($\Omega$) & 377 \\
        $Z_G$ ($\Omega$) & 377 \\
        $Z_0$ ($\Omega$) & 377/2.2 = 171 \\
        $\tau_0$ (ps) & 156 \\
        $\chi^{(2)}_{zzz}$ (pm/V) & 34.4\\
        
        \hline
    \end{tabular}
    \label{table:1}
\end{table}

\section*{References}
  \addcontentsline{toc}{section}{References}

{\small
\printbibliography[heading=none]
}

\noindent
\textbf{Supplementary information.}
Supplementary information includes:
Extended Figures 1-5
Supplementary derivations

\noindent
\textbf{Ackowledgements.} The authors acknowledge insightful discussions with Puneet Murthy and Edwin Ng, and thank Etienne Lorchat, Jim Wouda, Yassine Bakkouch, and Levin Seidt for their contributions to the experimental setup.  

\noindent
\textbf{Author contributions.}
O.K. and T.C. conceptualized the work. O.K., C.V., and T.C. fabricated the samples and assembled the experimental apparatus. O.K., C.V., and T.C. performed the measurements. O.K., C.V., and T.C. analyzed the results. O.K. derived the theoretical model. O.K. and T.C. wrote the manuscript. T.C. supervised the project.

\noindent
\textbf{Competing interests.} The authors declare no competing interests.

% \section*{References}
%   \addcontentsline{toc}{section}{References}

% {\small
% \printbibliography[heading=none]
% }

% \bmhead{Supplementary information}
% Supplementary information includes:
% Extended Figures 1-5
% Supplementary derivations

% \bmhead{Acknowledgments}
% The authors acknowledge insightful discussions with Puneet Murthy and Edwin Ng, and thank Etienne Lorchat, Jim Wouda, Yassine Bakkouch, and Levin Seidt for their contributions to the experimental setup.  

% \section*{Declarations}
% \textbf{Author contribution:}
% O.K. and T.C. conceptualized the work. O.K., C.V., and T.C. fabricated the samples and assembled the experimental apparatus. O.K., C.V., and T.C. performed the measurements. O.K., C.V., and T.C. analyzed the results. O.K. derived the theoretical model. O.K. and T.C. wrote the manuscript. T.C. supervised the project.

\end{document}